# Transition Temperatures of Superconductors estimated from Periodic Table Properties

**O. P. Isikaku-Ironkwe**[1, 2]
[1]The Center for Superconductivity Technologies (TCST)
Department of Physics,
Michael Okpara University of Agriculture, Umudike (MOUAU),
Umuahia, Abia State, Nigeria
[2]RTS Technologies, San Diego, CA 92122

## Abstract

**Predicting the transition temperature, Tc, of a superconductor from Periodic Table normal state properties is regarded as one of the grand challenges of superconductivity. By studying the correlations of Periodic Table properties with known superconductors, it is possible to estimate their transition temperatures. Starting from the isotope effect and correlations of superconductivity with electronegativity ($\mathcal{X}$), valence electron count per atom (Ne), atomic number(Z) and formula weight (Fw), we derive an empirical formula for estimating Tc that includes an unknown parameter,(Ko). With average values of $\mathcal{X}$, Ne and Z, we develop a material specific characterization dataset (MSCD) model of a superconductor that is quantitatively useful for characterizing and comparing superconductors. We show that for most superconductors, Ko correlates with Fw/Z, Ne, Z, number of atoms (An) in the formula, number of elements (En) and with Tc. We study some superconductor families and use the discovered correlations to predict similar and novel superconductors and also estimate their Tcs. Thus the material specific equations derived in this paper, the material specific characterization dataset (MSCD) system developed here and the discovered correlation between Tc and Fw/Z, En and An, provide the building blocks for the analysis, design and search of potential novel high temperature superconductors with specific estimated Tcs.**

## Introduction

**The transition temperature Tc, of a superconductor is one of its most important properties. Predicting the Tc of superconductors from Periodic Table [1] normal state properties is regarded as one of the challenges in the design and search for novel high temperature superconductors [2]. This search has been described as *"the very heart of scientific research in superconductivity"* [2]. Achieving Tc prediction has also been tagged a "grand challenge" that should lead to a paradigm shift from discovery by serendipity to discovery by design. The "*Basic Research Needs for Superconductivity*" report of 2006 [2, 3, 4, 5] observed that *"despite our greater experience with many more superconductors of very different kinds, however, we still do not know how to predict the occurrence of superconductivity or design a material with given superconducting properties".* The challenge of predicting superconducting properties of materials from the Periodic Table was first undertaken by B.T. Matthias [6, 7, 8, 9, 10] who observed a correlation of valence electron count per atom, Ne, with transition temperature, Tc. The isotope effect [11, 12] which pointed to a correlation between mass and Tc, became one of the building blocks of the BCS theory of superconductivity [13]. The BCS theory gave the first quantitative estimate of Tc. The BCS theory has however been criticized [14, 15] for not providing enough "material specific" parameters to search for and design superconductors. Villars and Phillips [16] had found three parameters (valence electrons, orbital radii and electronegativity) as "golden coordinates" that influence superconductivity and had high predictive power in their search for high-Tc superconductors. Hirsch recognized the grand challenge [17] and attempted a statistical approach to its solution by studying thirteen key variables, normal state properties that correlate with superconductivity. He proposed that it should be possible to find "normal state properties" that tell whether a material will be a superconductor or not. He found the Hall coefficient, $R_H$, to have the highest correlation factor followed by work function, bulk modulus, melting temperature, and atomic volume. Next in his list were Debye temperature, effective U, ionic mass, ionization potential, magnetic susceptibility and electrical conductivity. Specific heat and thermal conductivity came last. Many studies have found significant correlations between superconductivity and electronegativity [16, 18, 19, 20, 21, 22, 23], valence electrons [6, 7, 8,**

**9, 10, 24, 25, 26, 27] and atomic number [17]. Theoretical efforts such as the density functional theory, DFT, of superconductivity [28, 29] have made some initial success in solving the challenge, particularly with simple metals and alloys. DFT has however yet to incorporate the successful "enlightened" empirical approach [15] and "material specific" chemistry [14] aspects of discovering superconductivity. Our research has been influenced by the superconductivity report [2] and "The Hirsch question" [17], namely: *"Is it possible to accurately predict the magnitude of the superconducting critical temperature, Tc, of a material from measurements of its normal state properties?"***

**This paper attempts to answer the Hirsch question and to derive Tc as it relates to "material specific" Periodic Table properties such as electronegativity, valence electrons, atomic number, formula weight and other factors. These normal state properties are studied across many known families of superconductors and some correlations that may prove useful in designing and predicting the Tcs of novel families of superconductors are identified and quantified.**

**This paper is structured in three parts. Part I begins with the isotope effect. We then adopt the Gaussian distribution function to describe the maximum Tc to bound the variation of Tc with doping. We next extend it to correlations of the maximum Tc with Periodic Table properties such as electronegativity, valence electrons, formula weight and atomic number. For alloys and compounds we evaluate the average values of electronegativity, valence electron count and atomic number and the experimental parameter Ko at maximum Tc. In this work, the Pauling's Electronegativity Scale [18] and Matthias valence electron convention in reference 10 are used in Table 1. In Part I we also introduce a material-specific characterization dataset (MSCD) system for describing and characterizing superconductors.**

**In Part II we use these insights to explore material specific properties of known superconductors, expressing Ko in material specific terms for the families of superconductors studied. We extend the discovered correlations of Tc with formula weight over atomic number, Fw/Z, number of atoms, An, number of elements, En, and Ko to estimate the material specific properties of novel superconductors up to 400K.**

Part III discusses the implications of the material-specific paradigm introduced herein for the future analysis, design and search for potential high temperature superconductors (HTSCs). Guidelines for estimating superconductivity and Tc are enunciated, with examples.

## Part I: Basic Theoretical Framework

The Isotope effect [11, 12] provided the first evidence of a correlation between mass and transition temperature Tc, with Tc varying inversely as the square root of atomic mass (M) for many elements. Introducing a proportionality parameter K, Tc can be is expressed as:

$$T_c \approx \frac{K}{\sqrt{M}} \tag{1}$$

Not long after the discovery of the isotope effect, Matthias et al. [6 - 10] and others [24 - 27] showed that Tc is correlated to valence electrons per atom, Ne, in elements and intermetallic compounds. Experimental data [6, 8, 9, 25, 26, 27] suggests that the applicability of this correlation can be improved with the addition of a Gaussian function [30] and Tc can be expressed approximately as:

$$T_c \approx N_e \frac{1}{\sigma\sqrt{2\pi}}\, e^{-\frac{1}{2}y^2} \tag{2}$$

Where y = $\frac{(x-X)}{\sigma}$, in the Gaussian

$$G_{x,}(x) = \frac{1}{\sigma\sqrt{2\pi}}\, e^{-\frac{1}{2}y^2} \tag{3}$$

A similar Gaussian correlation also applies for variation of Tc with electronegativity, $\mathcal{X}$, [16, 17, 18, 19, 20] expressed approximately as:

$$T_c \approx \mathcal{X}\frac{1}{\sigma\sqrt{2\pi}}\, e^{-\frac{1}{2}y^2} \tag{4}$$

For this paper, we propose to combine the dependencies in equations 1, 2 and 4 to give:

$$T_c = \mathcal{X}\,\frac{Ne}{\sqrt{M}} K_1\, e^{-\frac{1}{2}y^2} \tag{5}$$

where $K_1$ is a modifying parameter to be determined by comparison with known superconductors. The atomic mass, M, is known to relate to its atomic number Z, by approximately:

$$M = K_2 Z \tag{6}$$

It can be shown that $K_1$ and $K_2$ can be combined and replaced by an appropriate parameter Ko, transforming equations (5) and (6) into:

$$T_c = \mathcal{X} \frac{\mathrm{Ne}}{\sqrt{Z}} K_o \; e^{-\frac{1}{2}y^2} \tag{7}$$

Note that equations 5 and 7 are similar except that the parameters $K_1$ and $K_0$ have different values. $K_1$ applies when we work with Mass and $K_o$ when we work with Atomic number. For the rest of this work, we will use $K_o$. $K_1$ and $K_o$ can be regarded as Tc modifying parameters unique for each material family. For example, Ko varies from 2.0 for the elements to about 70 for best known mercury –based high temperature superconductor. The limits to this equation are determined by the value of $e^{-1/2y2}$ since the other factors, except Tc, cannot be zero. For maximum Tc,

$$e^{-\frac{1}{2}y^2} = 1 \tag{8}$$

This occurs when y = 0. Equation 7 then reduces to:

$$T_c = \mathcal{X} \frac{\mathrm{Ne}}{\sqrt{Z}} K_o \tag{9}$$

Tc will be zero, when:

$$e^{-\frac{1}{2}y^2} = 0 \tag{10}$$

This occurs when y is infinite i.e. $y = \infty$. This value of y is determined by doping and crystal chemistry which we shall discuss in a future paper.

Though a Gaussian curve is assumed, the actual shape for any given superconductor will be determined by doping and yet to be determined chemical parameters that influence the value of y. The validity of the material specific equation 7 for Tc can be verified by studying **the combined effects on Tc of average values of electronegativity, $\mathcal{X}$, valence electrons, Ne, atomic number, Z, and the modifying parameter Ko. Next we derive the average values of these parameters.**

## Average Values of $\mathcal{X}$, Ne and Z

**Most superconducting materials are alloys or chemical compounds. According to Sanderson [31] the effective electronegativity $\mathcal{X}$ of a compound containing n elements is the average electronegativity of its constituent elements and atoms expressed as:**

$$\mathcal{X} = \frac{\sum_{i=1}^{n} \mathcal{X}iNi}{\sum_{i=1}^{n} Ni} \quad (11)$$

**where $N_i$ is the number of atoms of element i, with electronegativity $\mathcal{X}_i$. The same logic can be extended to averages of valence electron count, Ne, and atomic number, Z. Thus for a compound with formula $A_pB_qC_rD_sE_t$, where p, q, r, s, t represent the numbers of atoms of elements A, B, C, D and E respectively. In this work, we use electronegativity $\mathcal{X}$, as defined by Pauling [18] and Valence electron count per atom Ne, as defined by Matthias [10] and others [25, 26, 27]. The average values of $\mathcal{X}$, Ne and Z for multi-element systems can then be calculated as:**

$$\mathcal{X} = \frac{p\mathcal{X}_A + q\mathcal{X}_B + r\mathcal{X}_C + s\mathcal{X}_D + t\mathcal{X}_E}{p+q+r+s+t} \quad (12)$$

$$Ne = \frac{pNe_A + qNe_B + rNe_C + sNe_D + tNe_E}{p+q+r+s+t} \quad (13)$$

$$Z = \frac{pZ_A + qZ_B + rZ_C + sZ_D + tZ_E}{p+q+r+s+t} \quad (14)$$

**The total formula weight, Fw, of $A_pB_qC_rD_sE_t$ is expressed as:**

$$Fw = pFw_A + qFw_B + rFw_C + sFw_D + tFw_E \quad (15)$$

**The average values of electronegativity, valence electron count and atomic number will be used in computing the material specific characteristics of superconductors and other materials to be presented in this paper.**

## Material-Specific Characterization Dataset (MSCD)

**Using the Periodic Table[1] and Pauling's electronegativity scale (Table 1) we can compute from a material's formula, $A_pB_qC_rD_sE_t$ , its electronegativity, $\mathcal{X}$, valence electron count, Ne, atomic number, Z, and formula weight Fw, from equations 9, 12, 13, 14 and 15. When the Tc**

is known, Ko can be computed. We also compute the ratios: Ne/$\sqrt{Z}$, Fw/Z, Fw/Ne and Fw/$\sqrt{Z}$ for cases where Tc is not previously known. We shall show in Part II, how to compute Ko from correlations of Ko with some of the above parameters. Here we define the material specific characterization dataset (MSCD) of a superconductor $A_pB_qC_rD_sE_t$ as a set of parameters that chemically define the superconductor quantitatively.

$$A_pB_qC_rD_sE_t \equiv \langle \mathcal{X}, \text{Ne}, \text{Z}, \text{Fw}, \text{Tc}, \text{Ne}/\sqrt{Z}, K_0, \text{Fw/Z}, \text{Fw/Ne}, \text{An}, \text{En}, \text{Fw}/\sqrt{Z} \rangle \quad (16)$$

where An is the total number of atoms in the formula and En the total number of elements.

Besides been a powerful material specific computational characterization tool, the MSCD (16) is also useful for comparing superconductors. MSCD values computed using Table 1 and the Periodic Table of Elements [1] are displayed in Table 2 which represents the MSCD of many families of selected superconductors. MSCD may be seen as the "genome" of a superconductor in terms of the variables expressed therein. As we shall see in Part II of this paper, it can be used in the computational screening and combinatorial search for new superconductors.

# Part II: Analytical Observations of Superconductors

In this section we look for correlations of material specific properties that can be used for screening for superconductivity and also analyze the MSCD of families of superconductors and look for material-specific correlations of Tc with Fw/Z, An and En.

## 2.1 Computational Material-Specific Screening for HTSC

We found that the value of Ne/$\sqrt{z}$ provides a quick computational screening for superconductivity or its absence in many materials.
We found from MSCD in Table 2, that for most high-Tc superconductors,

$$0.8 < \text{Ne}/\sqrt{Z} < 1.0 \quad (17)$$

For the cuprates in particular, we observe that the electronegativity, $\mathcal{X}$, valence electron count, Ne, and atomic number, Z, lie within a narrow range specified as: $2.4 < \mathcal{X} < 2.6$; $3.8 < \text{Ne} < 4.4$ and $20 < Z < 25$.

## 2.2 Similar Superconductors

An interesting observation on superconductors with close enough Tcs is presented in the MSCD of Table 3 which lists five pairs of similar superconductors. We distinguish four possible cases:
(a) $\langle \mathcal{X}, \text{Ne} \rangle$ the same

**(b) ⟨Ne, Z⟩ the same**
**(c) ⟨$\mathcal{X}$, Z⟩ the same**
**(d) ⟨$\mathcal{X}$, Ne, Z⟩ the same.**
**We describe such superconductors as similar. We find the following empirical rule to apply within the range 0.75< Ne/$\sqrt{Z}$ < 1.02.**
**1. Materials with exactly the same average electronegativity, valence electrons and atomic number have the same transition temperature.**
**2. If two or more materials have the same average valence electrons Ne, and atomic number Z, then their Tcs will be proportional to their electronegativities.**
**3. If two or more materials have the same average electronegativity $\mathcal{X}$, and valence electrons Ne, then their Tcs will be proportional to their average atomic numbers, Z .**

## 2.3 Electronegativity

**Electronegativity[18] has been shown [19, 20, 21, 22, 23] to correlate with superconductivity. A low electronegativity ($\mathcal{X}<1.6$) is usually associated with low Tc. Tc tends to go up as electronegativity increases peaking between 2.5 and 2.6. for high temperature superconductors. Asokamani et al. [19] showed that pressure generally correlates with electronegativity: an increase in pressure leads to an increase in electronegativity. Most materials with electronegativity above 2.6 are not superconductors. Table 4 showing the variation of $\mathcal{X}$ with Tc for many superconductor families, indicates that electronegativity increases with Tc then at higher Tcs electronegativity tends to drop to around 2.46 (see figure 1).**

## 2.4 General Tc Correlations

**From the MSCD of Table 2 we produce Table 5 which represents the MSCD for six families of superconductors with 1, 2, 4 and 5 elements. We study the correlations of Tc with: Fw/Z, Ko, and En. These results are expressed as graphs in Figures 2, 3 and 4. Figure 2 shows that Tc increases with Fw/Z. Figure 3 also indicate that Tc increases with Ko. Fig. 4 shows that Tc increases with the number of elements.**

## 2.5 Material Specific Map for Superconductor Families

**Table 6 shows estimates for Ko in terms of Fw/Z for various families of superconductors, derived from Table 2 and references 25 and 26. Plugging these estimates into the formula for maximum Tc (equation 9), enables us to estimate the Tcs of families of superconductors indicated. We use these estimates in the next section to predict novel superconductors.**

## 2.6 Predicting New Superconductors & Estimating their TCs

**Using Table 6 and the rules in sections 2.1 and 2.2, we can predict novel superconductors as shown in Table 7 with 21 examples. We state their formulae and expected Tcs.**

### 2.7 Tc variation in the Hg Cuprates

The Mercury-based cuprates produce the highest Tc of all known superconductors. The origin of this high Tc in cuprates is still uncertain though detailed studies [32, 33, 40] attribute this high-Tc phenomenon to layered homologous systems. Graph of Tc versus number of layers for the mercury cuprates [32, 33] produce a very unique graph (figure 6) where Tc peaks at layer 3 at 133K, falling off on both sides. We have been able to reproduce this graph by plotting Tc versus Fw/Z for the mercury cuprates (figure 5). This strongly suggests that Fw/Z plays the same role as number of layers.

### 2.8 High Fw/Z: Origin of HTSC?

The significance of Fw/Z in high-Tc superconductivity is probed further by examining three superconductors [43, 44, 45] shown in Table 8. Tc moves from 40 K to 93K and to 133K as Fw/Z increases from slightly over 16 to 29 and then to over 36. Even though the number of elements also increased, from 3 to 5, the largest increase is in Fw/Z. This suggests that Fw/Z may be the most significant parameter in increasing Tc.

### 2.9 Estimating Parameters for HTSC higher than 133K

Using $HgBa_2Ca_2Cu_3O_8$ as a design model we assume $\frac{Ko}{Fw/z}$ =1.9, Ne = 3.85, An/En =3.2. With electronegativity $\mathcal{X}$, between 2.4 and 2.5 and $Ne/\sqrt{Z}$ around 0.8 and Ko values up to 200 we use these values to compute Tc in the formula: $T_c = \mathcal{X} \frac{Ne}{\sqrt{Z}} K_o$. Six examples are given with Tc almost 400K in Table 10. The computations also yield the formula weight and average atomic number of the expected superconductors in addition to the number of elements and atoms in the materials.

## Part III: Discussion, Conclusions & Acknowledgements

## Discussion

Estimating Tcs in material specific terms have been a fundamental unsolved problem [3, 4, 5, 14, 17, 39] --- till now. Inability to estimate Tc has led to failure in predicting new families of superconductors. This inability arose from the problem of identifying the limiting parameter or “glue” to high Tc[40, 41]. Our studies have identified it as Ko which in material specific terms is a function of Fw/Z , number of atoms, An, and number of elements En. This can be confirmed if we examine the change in Fw/Z with Tc as we go from Nb to $MgB_2$ to the Pnictides and cuprates (Table 2). There is a continuous large increase in Fw/Z, indicating a strong correlation with Tc as shown in Figure 2. Though the goal of this paper was not to delve into the mechanism of superconductivity, the data obtained showed that Fw/Z is the deciding parameter or “The Glue” to high temperature superconductivity. For each family of HTSC, Tc peaks at some value of Fw/Z which is not the highest value of Fw/Z for that family. The optimum Fw/Z for highest Tc is yet to be explained. We however found an empirical relation between maximum Tc and number of atoms An and elements En. An/En = 3.2 for $HgBa_2Ca_2Cu_3O_8$, our model design example. Tc definitely increases with Fw/Z, number of **elements and atoms then peaks and descends for each family of superconductors,(see figure 5 for Mercury cuprates).**

**In this study, we did not use the traditional tools of DFT and band structure calculations. Treating superconductivity as a material-specific chemistry problem, we rather explored chemical correlations and the power of the Periodic Table of elements [46]. Our predictions of 21 superconductors based on this novel approach, is the implicit proof of its correctness or not.**

**The techniques developed here have not directly taken into account structural features, phases, pressure and doping which really are embedded in electronegativity [19] and valence electrons [27] and atomic number. Nor have we delved into the theories of superconductivity to derive the equations used. We have used the minimal assumptions on the chemical foundations of superconductivity. We recognize the millions of possible permutations and combinations of the elements that could yield superconductivity [34]. However the methods introduced here present a novel computational combinatorial paradigm very different from previously proposed methods [36, 37]. It also considerably reduces the search space[34].**

**The magnetic features of superconductivity and the physics of heavy fermion superconductivity [51, 52] are outside the scope of this preliminary study focused on high temperature superconductors. They will be treated in a future paper.**

**Many general guidelines have been advanced for searching for high-Tc superconductors [15, 42, 46, 47, 48, 49, 50]. The guidelines we have in this paper are unique in that they are material specific and actually estimate the Tcs you will get if you use them and the chemical formula of the superconductors that give the Tcs. We have achieved this with simple concepts of chemistry and physics using correlations of electronegativity, valence electron count, atomic number, formula weight over Z with superconductivity.**

## Conclusions

**The rational design of superconductors depends on obtaining a reasonable estimate for Tc which depends on identifying the parameters that influence Tc. In this paper we have identified the parameters as averages of electronegativity, valence electrons count per atom, formula weight over atomic number, Fw/Z, number of atoms, An, and number of elements En, in the material and a parameter Ko(equation 9). We have derived an empirical formula for Tc based on these parameters.**

**We found a material specific equivalent of Ko as a weighted value of formula weight divided by average atomic number of the constituent elements (Fw/Z). The value of Fw/Z is a key determinant in the value of Tc. We showed that from the Fw/Z of a superconductor and the number of elements and atoms we can tell the limits of its Tc.**

**We also established material specific tests for HTSC over 90K (0.75< Ne/$\sqrt{Z}$ <1.0); 2.4<$\mathcal{X}$<2.6 ; 3.8<Ne<4.4; 20<Z<25. We also proposed three empirical rules for detecting similar superconductors, namely:**

**1. Materials with exactly the same average electronegativity, valence electrons and atomic number have the same transition temperature.**

**2. If two or more materials have the same average valence electrons Ne, and atomic number Z, then their Tcs will be proportional to their electronegativities.**

**3. If two or more materials have the same average electronegativity and valence electrons, then their Tcs will be proportional to their average atomic numbers.**

**Using these rules and the material specific equations 9, 11 – 14, we predict 21 new superconductors and their Tc, displayed in Table 7.**

**We estimated in Table 10, material-specific parameters required to achieve room temperature superconductivity [15, 36, 37, 38, 39, 41, 42, 45, 47]. Thus we can assert that we have successfully taken on the grand challenge and answered in the affirmative the Hirsch question: "*Is it possible to accurately predict the magnitude of the superconducting critical temperature, Tc, of a material from measurements of its normal state properties?*" Finally we propose that the methodology developed in this paper provide a novel set of building blocks for analysis, design and search for high temperature superconductors with Tc up to 400K.**

## Acknowledgements

**It is a pleasure to acknowledge stimulating discussions with A.O.E. Animalu at University of Nigeria, Nsukka. Coming to UC San Diego in 2011, M.B. Maple, J.E. Hirsch, and J. Fortin asked critical questions that helped shape and focus the research. M. J. Schaffer, then at General Atomics, San Diego was always available to discuss, correct and suggest better ways to present the concepts and assumptions in the paper. J.R. O'Brien at Quantum Design provided me with much needed research resources in addition to stimulating discussions on chemistry aspects of superconductivity. This research was financially supported during this period by M. J. Schaffer.**

# LIST OF TABLES AND FIGURES

List of Tables

**Data for the Tables are obtained from References 25, 26, 35 and 45. Computations for Ko are done with equation (9) for maximum Tc:** $\mathbf{T_c} = \mathcal{X}\frac{\mathbf{Ne}}{\sqrt{Z}}\mathrm{K_o}$

## List of Figures

**Figure 1: Tc vs $\mathcal{X}$ for 6 families of superconductors**

**Figure 2: Tc vs Fw/Z for 6 families of superconductors**

**Figure 3: Tc vs Ko for 6 families of superconductors**

**Figure 4: Tc vs En for 6 families of superconductors**

**Figure 5: Tc vs Fw/Z for Mercury Cuprates**

**Figure 6: Tc vs No. of layers for Mercury Cuprates**

| Ne = 1 | | Ne = 2 | | Ne = 3 | | Ne = 4 | | Ne = 5 | | Ne = 6 | |
|---|---|---|---|---|---|---|---|---|---|---|---|
| E | $\mathcal{X}$ | E | $\mathcal{X}$ | E | $\mathcal{X}$ | E | $\mathcal{X}$ | E | $\mathcal{X}$ | E | $\mathcal{X}$ |
| H | 2.1 | Be | 1.5 | Sc | 1.3 | Ti | 1.5 | V | 1.6 | Cr | 1.6 |
| Li | 1.0 | Mg | 1.2 | Y | 1.2 | Zr | 1.4 | Nb | 1.6 | Mo | 1.8 |
| K | 0.8 | Ca | 1.0 | La | 1.1 | Hf | 1.3 | Ta | 1.5 | W | 1.7 |
| Rb | 0.8 | Sr | 1.0 | B | 2.0 | C | 2.5 | N | 3.0 | O | 3.5 |
| Cs | 0.7 | Ba | 0.9 | Al | 1.5 | Si | 1.8 | P | 2.1 | S | 2.5 |
| Cu | 1.9 | Zn | 1.6 | Ga | 1.6 | Ge | 1.8 | As | 2.0 | Se | 2.4 |
| Ag | 1.9 | Cd | 1.7 | In | 1.7 | Sn | 1.8 | Sb | 1.9 | Te | 2.1 |
| | | Hg | 1.9 | Tl | 1.8 | Pb | 1.8 | Bi | 1.9 | | |

| Ne = 7 | | Ne = 8 | | Ne = 9 | | Ne = 10 | | |
|---|---|---|---|---|---|---|---|---|
| E | $\mathcal{X}$ | E | $\mathcal{X}$ | E | $\mathcal{X}$ | E | $\mathcal{X}$ | **Nomenclature** |
| Mn | 1.5 | Fe | 1.8 | Co | 1.8 | Ni | 1.8 | $\mathcal{X}$ = Electronegativity |
| Tc | 1.9 | Ru | 2.2 | Rh | 2.2 | Pd | 2.2 | E = Element |
| Re | 1.9 | Os | 2.2 | Ir | 2.2 | Pt | 2.2 | Ne = Valence Electron Count |
| F | 4.0 | | | | | | | |
| Cl | 3.0 | | | | | | | |
| Br | 2.8 | | | | | | | |
| I | 2.5 | | | | | | | |

| Ne = 3 | | | | | | | | | | | | | | | | |
|---|---|---|---|---|---|---|---|---|---|---|---|---|---|---|---|---|
| E | Ce | Pr | Nd | Pm | Eu | Tb | Yb | Lu | Sm | Gd | Ho | Er | Tm | Th | Pu | U |
| $\mathcal{X}$ | 1.1 | 1.1 | 1.1 | 1.1 | 1.1 | 1.1 | 1.1 | 1.1 | 1.2 | 1.2 | 1.2 | 1.2 | 1.3 | 1.3 | 1.3 | 1.4 |

**Table 1: Pauling's Electronegativity Scale and Matthias's Valence Electron Count Convention**

| | Superconductor | $\mathcal{X}$ | Ne | Z | $Ne/\sqrt{Z}$ | Fw | Fw/Z | Tc(K) | Ko | An | En |
|---|---|---|---|---|---|---|---|---|---|---|---|
| 1 | Nb | 1.60 | 5.0 | 41 | 0.7809 | 92.91 | 2.27 | 9.2 | 7.14 | 1 | 1 |
| 2 | NbN | 2.30 | 5.0 | 24 | 1.0206 | 106.913 | 4.45 | 17 | 7.55 | 2 | 2 |
| 3 | MoC | 2.15 | 5.0 | 24 | 1.0206 | 107.951 | 4.5 | 14.3 | 6.79 | 2 | 2 |
| 4 | $Nb_3Ge$ | 1.65 | 4.75 | 38.75 | 0.7631 | 351.34 | 9.07 | 23.2 | 18.43 | 4 | 2 |
| 5 | $Y_2C_3$ | 1.98 | 3.6 | 19.2 | .8216 | 213.85 | 11.14 | 18 | 11.07 | 5 | 2 |
| 6 | $MgB_2$ | 1.7333 | 2.667 | 7.333 | 0.9847 | 45.93 | 6.263 | 39 | 22.85 | 3 | 2 |
| 7 | $Ba_{0.6}K_{0.4}BiO_3$ | 2.652 | 4.92 | 29.64 | 0.9037 | 355.02 | 11.98 | 30 | 12.52 | 5 | 3 |
| 8 | $La_{1.85}Sr_{.15}CuO_4$ | 2.5836 | 4.4071 | 24.593 | 0.8905 | 397.68 | 16.17 | 40 | 17.39 | 7 | 3 |
| 9 | $Ba_{0.6}K_{0.4}Fe_2As_2$ | 1.692 | 5.52 | 31.84 | 0.9783 | 359.58 | 11.293 | 38 | 22.96 | 5 | 3 |
| 10 | $Sr_{0.6}K_{0.4}Fe_2As_2$ | 1.704 | 5.52 | 29.68 | 1.0132 | 329.752 | 11.110 | 37 | 21.43 | 5 | 3 |
| 11 | $Ba_{0.6}K_{0.4}PbO_3$ | 2.632 | 4.72 | 29.44 | 0.8699 | 353.24 | 11.999 | 30 | 13.10 | 5 | 3 |
| 12 | $SmFeAsO_{0.9}F_{0.1}$ | 2.1375 | 5.525 | 32.275 | 0.9725 | 296.43 | 9.185 | 55 | 26.46 | 4 | 4 |
| 13 | $Gd_{0.8}Th_{0.2}FeAsO$ | 2.11 | 5.5 | 34.05 | 0.9125 | 318.98 | 9.37 | 56.3 | 29.24 | 4 | 4 |
| 14 | $GdFeAsO_{0.85}$ | 2.0455 | 5.481 | 32.45 | 0.9621 | 301.62 | 9.295 | 53.5 | 27.19 | 4 | 4 |
| 15 | $YBa_2Cu_3O_7$ | 2.5538 | 4.0 | 22.615 | 0.8411 | 666.22 | 29.46 | 93 | 43.3 | 13 | 4 |
| 16 | $YBa_2Cu_4O_8$ | 2.5733 | 3.933 | 22.067 | 0.8373 | 745.77 | 33.80 | 80 | 37.13 | 15 | 4 |
| 17 | $Tl_2Ba_2CuO_6$ | 2.5727 | 4.273 | 31.909 | 0.7564 | 842.97 | 26.42 | 80 | 41.11 | 11 | 4 |
| 18 | $Tl_2Ba_2CaCu_2O_8$ | 2.5467 | 4.133 | 27.733 | 0.7849 | 978.56 | 35.29 | 108 | 54.03 | 15 | 5 |
| 19 | $Tl_2Ba_2Ca_2Cu_3O_{10}$ | 2.5316 | 4.053 | 25.316 | 0.8054 | 1114.23 | 44.01 | 125 | 61.30 | 20 | 5 |
| 20 | $Bi_2Sr_2CuO_6$ | 2.6091 | 4.636 | 29.0 | 0.8609 | 752.75 | 25.96 | 22 | 9.79 | 11 | 4 |
| 21 | $Bi_2Sr_2\ CaCu_2O_8$ | 2.5733 | 4.4 | 25.6 | 0.8696 | 888.38 | 34.702 | 92 | 41.11 | 15 | 5 |
| 22 | $Bi_2Sr_2\ Ca_2Cu_3O_{10}$ | 2.5526 | 4.263 | 23.632 | 0.8770 | 1024.01 | 43.332 | 110 | 49.12 | 19 | 5 |
| 23 | $TlBa_2CuO_5$ | 2.5556 | 4.222 | 29.111 | 0.7825 | 622.59 | 21.39 | 52 | 26.00 | 9 | 4 |
| 24 | $Pb_2Ca_3Cu_2O_8$ | 2.56 | 4.267 | 23.007 | 0.8884 | 789.73 | 34.24 | 110 | 48.37 | 15 | 4 |
| 25 | $HgBa_2CuO_4$ | 2.45 | 3.875 | 31.625 | 0.6891 | 602.80 | 19.06 | 95 | 56.27 | 8 | 4 |
| 26 | $HgBa_2CaCu_2O_6$ | 2.4583 | 3.833 | 26.50 | 0.7446 | 738.43 | 27.87 | 122 | 66.65 | 12 | 5 |
| 27 | $HgBa_2\ Ca_2Cu_3O_8$ | 2.4625 | 3.813 | 23.938 | 0.7792 | 874.044 | 36.51 | 133 | 69.32 | 16 | 5 |

**Table 2: Material Specific Characterization Dataset (MSCD) for 27 Superconductors with Tcs between 9K and 133K**

| | Superconductor | $\mathcal{X}$ | Ne | Z | $Ne/\sqrt{Z}$ | Fw | Fw/Z | Tc(K) | Ko | An | En |
|---|---|---|---|---|---|---|---|---|---|---|---|
| 1 | NbN | 2.30 | 5.0 | 24 | 1.0206 | 106.913 | 4.45 | 17 | 7.55 | 2 | 2 |
| | MoC | 2.15 | 5.0 | 24 | 1.0206 | 107.951 | 4.50 | 14.3 | 6.79 | 2 | 2 |
| 2 | CaAlSi | 1.4333 | 3.0 | 15.667 | .7579 | 95.15 | 6.07 | 7.8 | 7.18 | 3 | 3 |
| | SrAlSi | 1.4333 | 3.0 | 21.667 | .6445 | 142.69 | 6.59 | 5.8 | 6.28 | 3 | 3 |
| 3 | ZrN | 2.2 | 4.5 | 23.5 | .9283 | 105.231 | 4.478 | 10.7 | 5.24 | 2 | 2 |
| | NbC | 2.05 | 4.5 | 23.5 | .9283 | 104.917 | 4.465 | 11.1 | 5.83 | 2 | 2 |
| 4 | $Nb_3Sn$ | 1.65 | 4.75 | 43.25 | .7181 | 397.44 | 9.19 | 18 | 15.19 | 4 | 2 |
| | $Nb_3Ge$ | 1.65 | 4.75 | 38.75 | .7631 | 351.34 | 9.07 | 23.2 | 18.43 | 4 | 2 |
| 5 | $BeB_2$ | 1.8333 | 2.6667 | 4.6667 | 1.2344 | 30.63 | 6.56 | 0 | 0 | 3 | 2 |
| | LiBC | 1.8333 | 2.6667 | 4.6667 | 1.2344 | 29.96 | 6.42 | 0 | 0 | 3 | 3 |

**Table 3: MSCD of similar superconductors with close Tcs. Five pairs cases of similar superconductors with: (a)$\langle \mathcal{X}, Ne \rangle$ the same: pairs 2, 4 and 5; (b) $\langle Ne, Z \rangle$ the same: pairs 1, 3 and 5. (c) $\langle \mathcal{X}, Ne, Z \rangle$ the same; pair 5. Note the closeness of their Tcs. In case 5 they have exactly the same Tc.**

| | $\mathcal{X}$ | Tc (K) | Material |
|---|---|---|---|
| 1 | 1.73 | 39 | $MgB_2$ |
| 2 | 2.11 | 56 | $Gd_{0.8}Th_{0.2}FeAsO$ |
| 3 | 2.57 | 80 | $Tl_2Ba_2CuO_6$ |
| 4 | 2.56 | 110 | $Bi_2Sr_2 Ca_2Cu_3O_{10}$ |
| 5 | 2.53 | 125 | $Tl_2Ba_2Ca_2Cu_3O_{10}$ |
| 6 | 2.46 | 133 | $HgBa_2 Ca_2Cu_3O_8$ |

**Table 4: Tc vs Electronegativity for 6 families of superconductors.**

| Example | $\mathcal{X}$ | Ne | Z | $Ne/\sqrt{Z}$ | Fw/Z | Ko | En | An | Tc (K) |
|---|---|---|---|---|---|---|---|---|---|
| **Nb** | **1.6** | **5.0** | **41.0** | **0.781** | **2.27** | **7.14** | **1** | **1** | **9.2** |
| **NbN** | **2.3** | **5.0** | **24.0** | **1.0206** | **4.45** | **7.55** | **2** | **2** | **17** |
| **$MgB_2$** | **1.73** | **2.667** | **7.333** | **0.985** | **6.26** | **22.85** | **2** | **3** | **39** |
| **$Gd_{0.8}Th_{0.2}FeAsO$** | **2.11** | **5.5** | **34.05** | **0.9125** | **9.37** | **29.24** | **4** | **4** | **56** |
| **$Pb_2Ca_3Cu_2O_8$** | **2.56** | **4.267** | **23.01** | **0.888** | **34.24** | **48.37** | **4** | **15** | **110** |
| **$HgBa_2 Ca_2Cu_3O_8$** | **2.463** | **3.813** | **23.94** | **0.779** | **36.51** | **69.32** | **5** | **16** | **133** |

**Table 5: MSCD for 6 Families of superconductors with En = 1, 2, 4 and 5 at ambient pressure**

| | Family | Ko Estimate | Comments |
|---|---|---|---|
| 1 | $AB_2$ | 3.65(Fw/Z) | Applies to $MgB_2$ only |
| 2 | $A_2B_3$, $ABC_3$, $AB_4$ | Fw/Z | Applies to $Y_2C_3$, $Ba_{0.6}K_{0.4}BiO_3$, and similar materials. |
| 3 | $AB_2C_2$ | 2(Fw/Z) | Applies to Pnictide 122 compounds |
| 4 | Pnictide 1-1-1-1 | $0.25(Fw/Ne + Fw/\sqrt{Z})$ | Applies at Tc(max) |
| 5 | $YBa_2Cu_3O_7$ | Fw/Z + An | An = number of atoms in formula |
| 6 | $Pb_2Ca_3Cu_2O_8$ | Fw/Z + An | An = number of atoms in formula |
| 7 | $HgBa_2 Ca_2Cu_3O_8$ | 1.9(Fw/Z) | Applies to Hg cuprates at highest Tc |

**Table 6: Material-specific estimates for Ko in terms of Fw, Ne, Z, An and En for families of superconductors in the formula:** $T_c = \mathcal{X}\frac{Ne}{\sqrt{Z}}K_o$

| | Model | Predicted Materials | | Predicted Tc Range |
|---|---|---|---|---|
| A | $MgB_2$ (Tc =39K) | LiMgN | 1 | 39K |
| | | $Li_2S$ | 2 | 30K |
| | | $Mg_2C$ | 3 | 31.5K |
| | | MgBeC | 4 | 39K |
| | | KCN | 5 | 20K |
| | | NaBC | 6 | 29K |
| | | $KB_5$ | 7 | 39K |
| | | $NbSi_2$ | 8 | 25 - 30K |
| | | $Li_2Mg_2CO$ | 9 | 35.8K |
| | | $Na_2Be_2CO$ | 10 | 35.8K |
| | | $Na_2Be_2CS$ | 11 | 35.1K |
| B | NbN(Tc=17K) MoC(Tc=14.3K) | GeS | 12 | 14.3 - 17K |
| | | SiSe | 13 | 14.3 - 17K |
| C | LiBC(Tc=0K) | $Be_2C$ | 14 | 0K |
| | | $Na_2O$ | 15 | 0K |
| D | CaBeSi (Tc=0.4K) | $Na_2S$ | 16 | 0.4K |
| E | $Y_2C_3$ (Tc=18K) | $Y_2S_3$ | 17 | 20.6K |
| | | $In_2S_3$ | 18 | 21.6K |
| | | $K_2S_3$ | 19 | 17.8K |
| | | $Al_2Se_3$ | 20 | 22K |
| | | $As_2Se_3$ | 21 | 24.9K |

**Table 7: Some 21 predicted material specific superconductors and their Tcs based on equation (9), Table 6 and the similarity rules proposed in sections 2.1 and 2.2.**

| | Superconductor | $\mathcal{X}$ | Ne | Z | Ne/$\sqrt{Z}$ | Fw | Fw/Z | Tc(K) | Ko | An | En |
|---|---|---|---|---|---|---|---|---|---|---|---|
| 1 | $La_{1.85}Sr_{.15}CuO_4$ | 2.5836 | 4.4071 | 24.593 | 0.8905 | 397.68 | 16.17 | 40 | 17.39 | 7 | 3 |
| 2 | $YBa_2Cu_3O_7$ | 2.5538 | 4.0 | 22.615 | 0.8411 | 666.22 | 29.46 | 93 | 43.3 | 13 | 4 |
| 3 | $HgBa_2\ Ca_2Cu_3O_8$ | 2.4625 | 3.8125 | 23.938 | 0.7792 | 874.044 | 36.51 | 133 | 69.32 | 16 | 5 |

**Table 8: Significance of Fw/Z in raising Tc from 40K to 93K and to 133K. Fw/Z may be "the glue" to achieving high-Tc superconductivity. Note too that the number of elements and atoms increased with Tc. The product {$\mathcal{X}$. Ne/$\sqrt{Z}$} did not change as significantly as Fw/Z.**

| | Superconductor | $\mathcal{X}$ | Ne | Z | Ne/$\sqrt{Z}$ | Fw | Fw/Z | Tc [K] | $K_0$ | $\frac{Ko}{Fw/z}$ | An | En |
|---|---|---|---|---|---|---|---|---|---|---|---|---|
| 1 | $HgBa_2CuO_4$ | 2.45 | 3.875 | 31.625 | 0.6891 | 602.8 | 19.06 | 97 | 57.45 | 3.02 | 8 | 4 |
| 2 | $HgBa_2CaCu_2O_6$ | 2.4583 | 3.8333 | 26.5 | 0.7446 | 736.43 | 27.79 | 126 | 68.83 | 2.48 | 12 | 5 |
| 3 | $HgBa_2Ca_2Cu_3O_8$ | 2.4625 | 3. 8125 | 23.938 | .7792 | 874.06 | 36.51 | 133 | 70.46 | 1.93 | 16 | 5 |
| 4 | $HgBa_2Ca_3Cu_4O_{10}$ | 2.465 | 3.80 | 22.4 | 0.8029 | 1009.69 | 45.08 | 125 | 63.16 | 1.40 | 20 | 5 |
| 5 | $HgBa_2Ca_4Cu_5O_{12}$ | 2.4667 | 3.7917 | 21.375 | 0.8201 | 1145.32 | 53.58 | 110 | 54.38 | 1.02 | 24 | 5 |
| 6 | $HgBa_2Ca_5Cu_6O_{14}$ | 2.4679 | 3.7857 | 20.643 | 0.8332 | 1280.95 | 62.05 | 97 | 47.17 | 0.76 | 28 | 5 |
| 7 | $HgBa_2Ca_6Cu_7O_{16}$ | 2.4688 | 3.7813 | 20.094 | 0.8435 | 1416.58 | 70.50 | 88 | 42.26 | 0.6 | 32 | 5 |

**Table 9: Hg-based Cuprate superconductors. Data from Ref. 26.**

| | Superconductor | $\mathcal{X}$ | Ne/$\sqrt{Z}$ | $K_0$ | Tc [K] | Fw/Z | $\frac{Ko}{Fw/z}$ | Fw | Ne | Z | An | En |
|---|---|---|---|---|---|---|---|---|---|---|---|---|
| 1 | **$HgBa_2Ca_2Cu_3O_8$** | 2.46 | .7792 | 69.32 | 133 | 36.51 | 1.93 | 874.1 | 3.81 | 23.94 | 16 | 5 |
| 2 | **HTSC-A1** | 2.5 | 0.8 | 100 | 200 | 52.63 | 1.9 | 1218.96 | 3.85 | 23.16 | 23 | 7 |
| 3 | **HTSC-A2** | 2.45 | 0.8 | 120 | 235.2 | 63.16 | 1.9 | 1462.79 | 3.85 | 23.16 | 28 | 9 |
| 4 | **HTSC-A3** | 2.5 | 0.8 | 140 | 280 | 73.68 | 1.9 | 1706.43 | 3.85 | 23.16 | 33 | 10 |
| 5 | **HTSC-A4** | 2.45 | 0.79 | 160 | 309.7 | 82.21 | 1.9 | 1952.50 | 3.85 | 23.75 | 36 | 11 |
| 6 | **HTSC-A6** | 2.45 | 0.8 | 180 | 352.8 | 94.74 | 1.9 | 2194.19 | 3.85 | 23.16 | 42 | 13 |
| 7 | **HTSC-A6** | 2.55 | 0.78 | 200 | 397.8 | 105.26 | 1.9 | 2564.46 | 3.85 | 24.36 | 47 | 15 |

**Table 10:** Computations for possible ambient pressure high-Tc Superconductors (HTSC-Ax) to over 390K, assuming some initial values for $\mathcal{X}$, Ne/$\sqrt{Z}$ and Ko in the formula: **Tc =** $\mathcal{X} \frac{Ne}{\sqrt{Z}} K_o$.
Assuming a $\frac{Ko}{Fw/z}$ =1.9, we compute Fw/Z for the proposed HTSC-Ax materials and also the Fw and Z, assuming Ne = 3.85. An =number of atoms and En =number of elements in the composition, assuming **An/En =3.2** , like in **$HgBa_2Ca_2Cu_3O_8$** and **$YBa_2Cu_3O_7$**

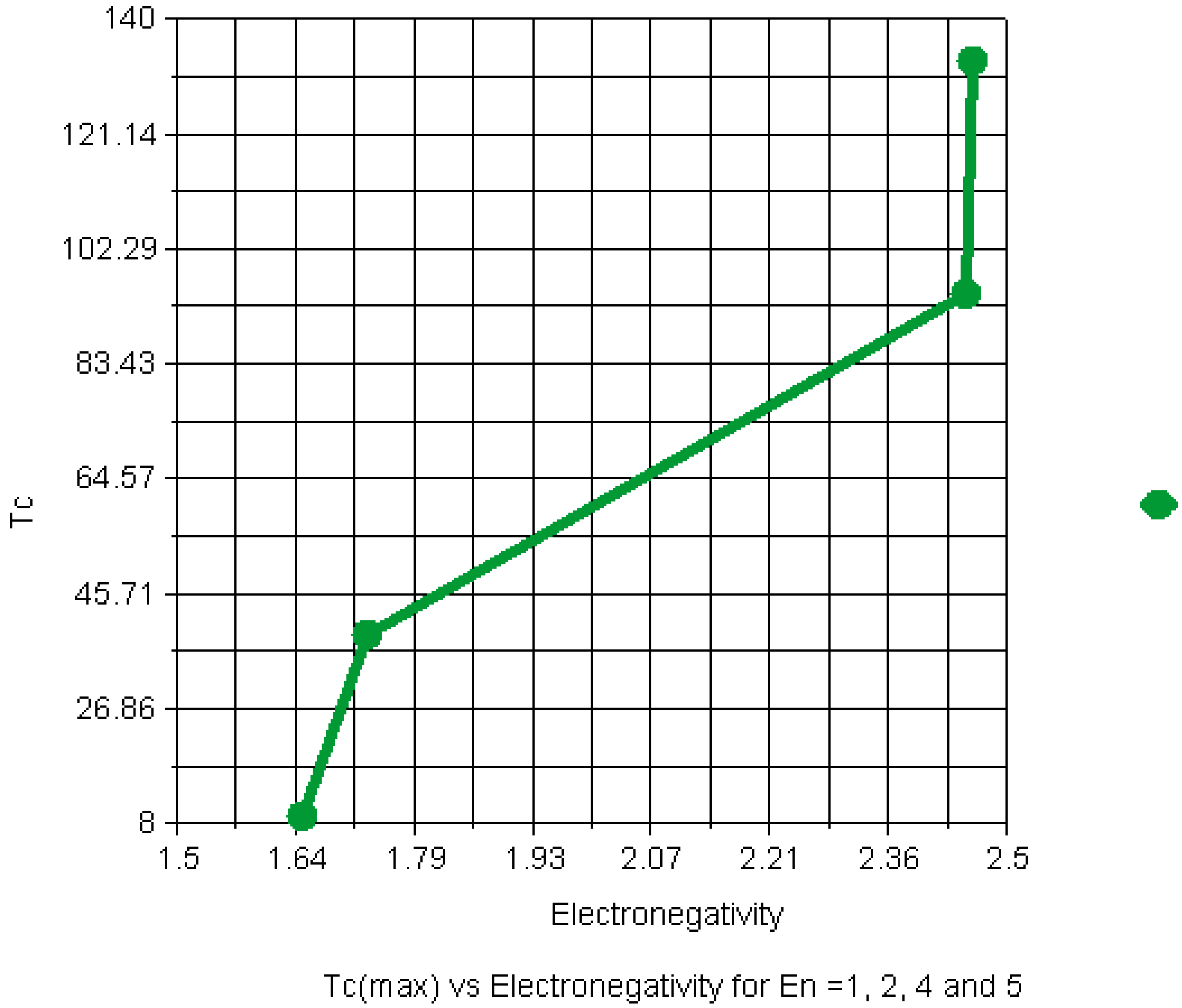

**Figure 1: Tc (max) vs Electronegativity for En = 1, 2, 4 and 5. We observe there is an electronegativity range between 1.73 and 2.4 which should have novel superconductors with Tc between 39K and 90K. So far only the Pnictides (Tc = 56K) have been discovered in this range. Between 2.4 and 2.5, Tc increases without much increase in electronegativity.**

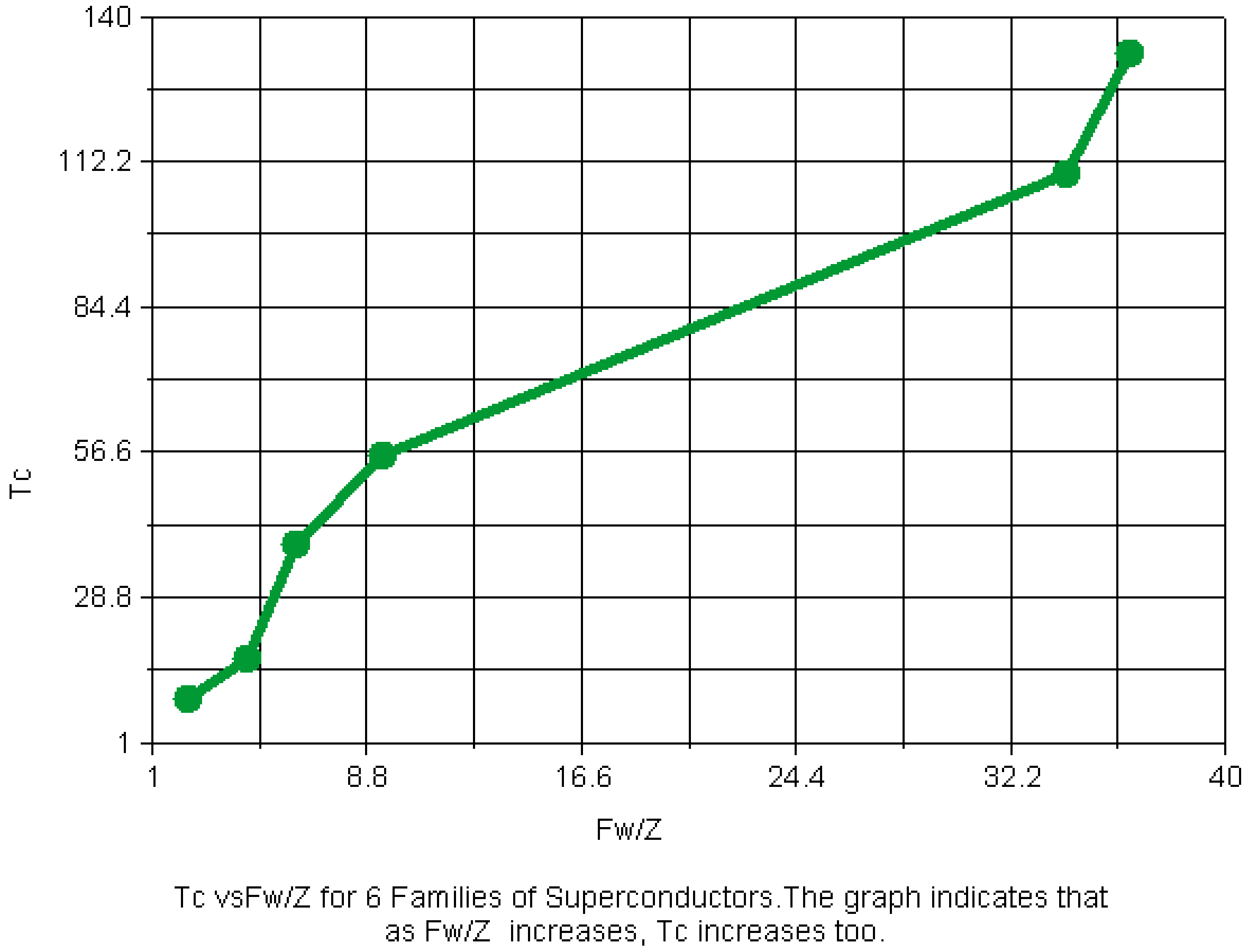

**Figure 2: Tc (max) vs Fw/Z for for 6 Families of superconductors with En =1, 2, 4 and 5 using data of Table 5. From the graph, we can see that there are yet to be discovered superconductor families with Tc between 60 and 100K and with Fw/Z between 10 and 30. The En may be 3 or 4.**

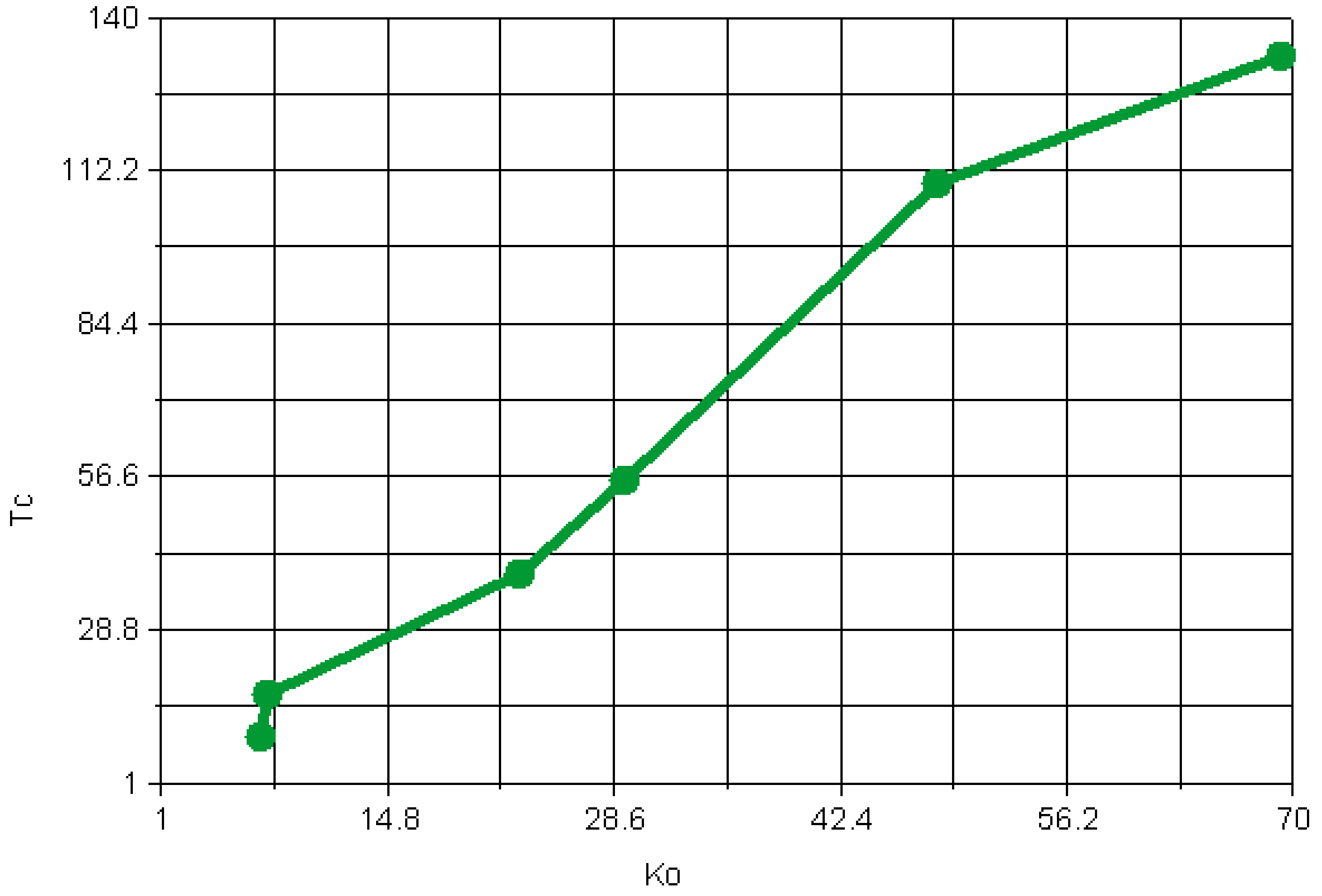

Fig. 2: Tc vs Ko for 6 Families of Superconductors.The graph indicates that as Ko increases, Tc increases too.

**Figure 3: Tc vs Ko for 6 Families of superconductors. The graph shows undiscovered superconductors with Ko between 30 and 42 and with Tc between 60K and 90K.**

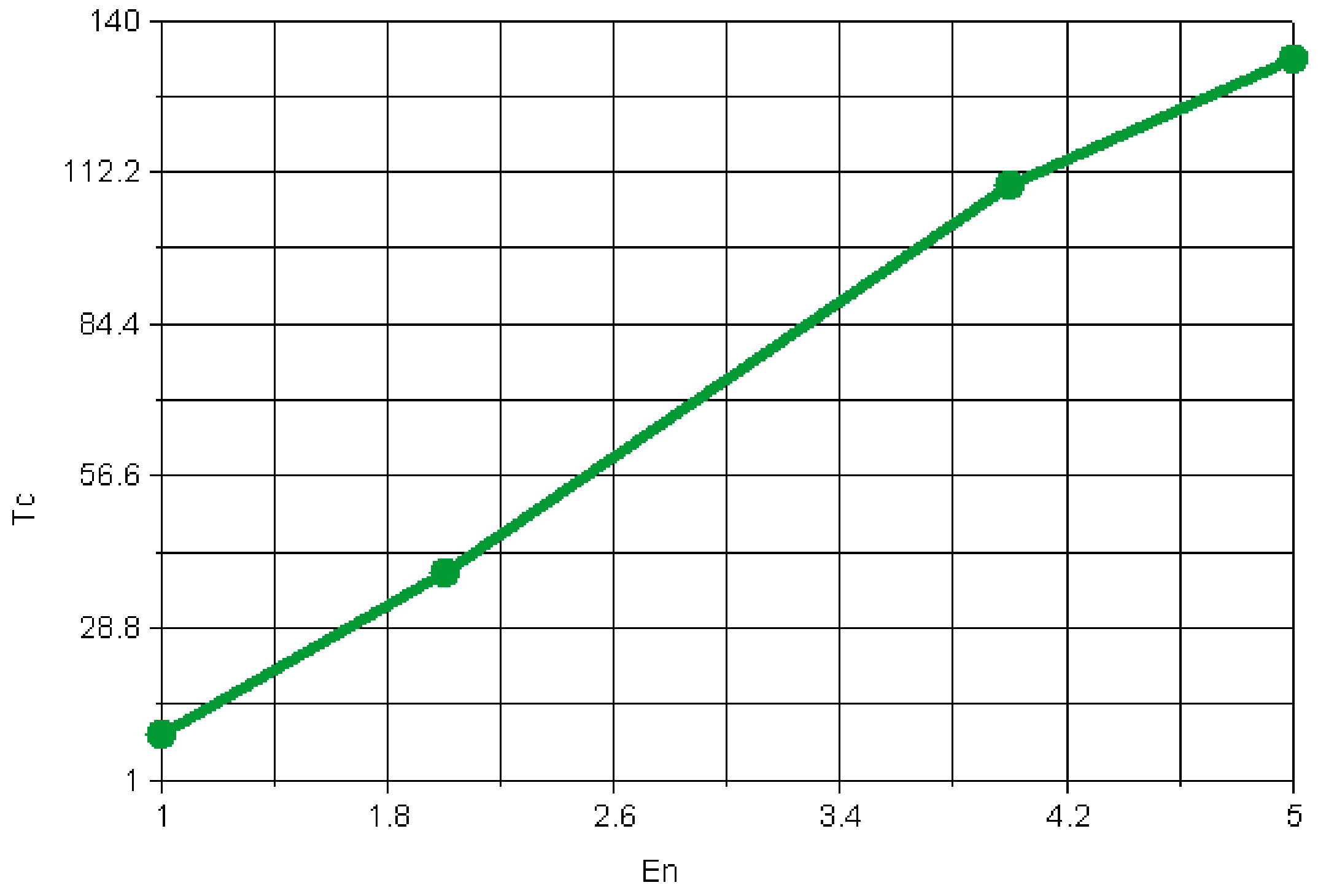

Fig. 4: Tc vs En for 6 Families of Superconductors.The graph indicates that as En increases, Tc increases too.

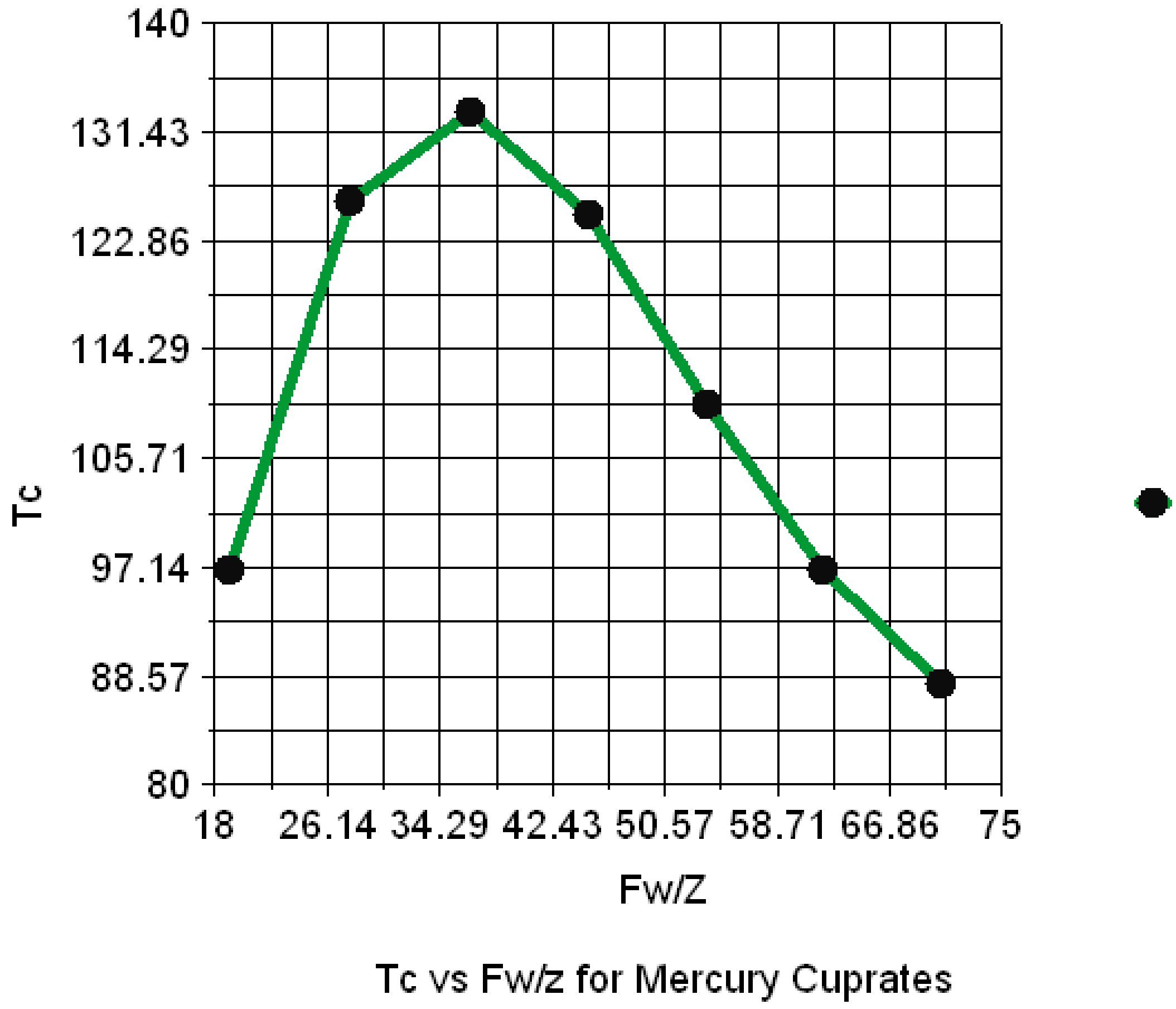

**Figure 5: Tc vs Fw/Z for Mercury Cuprates**

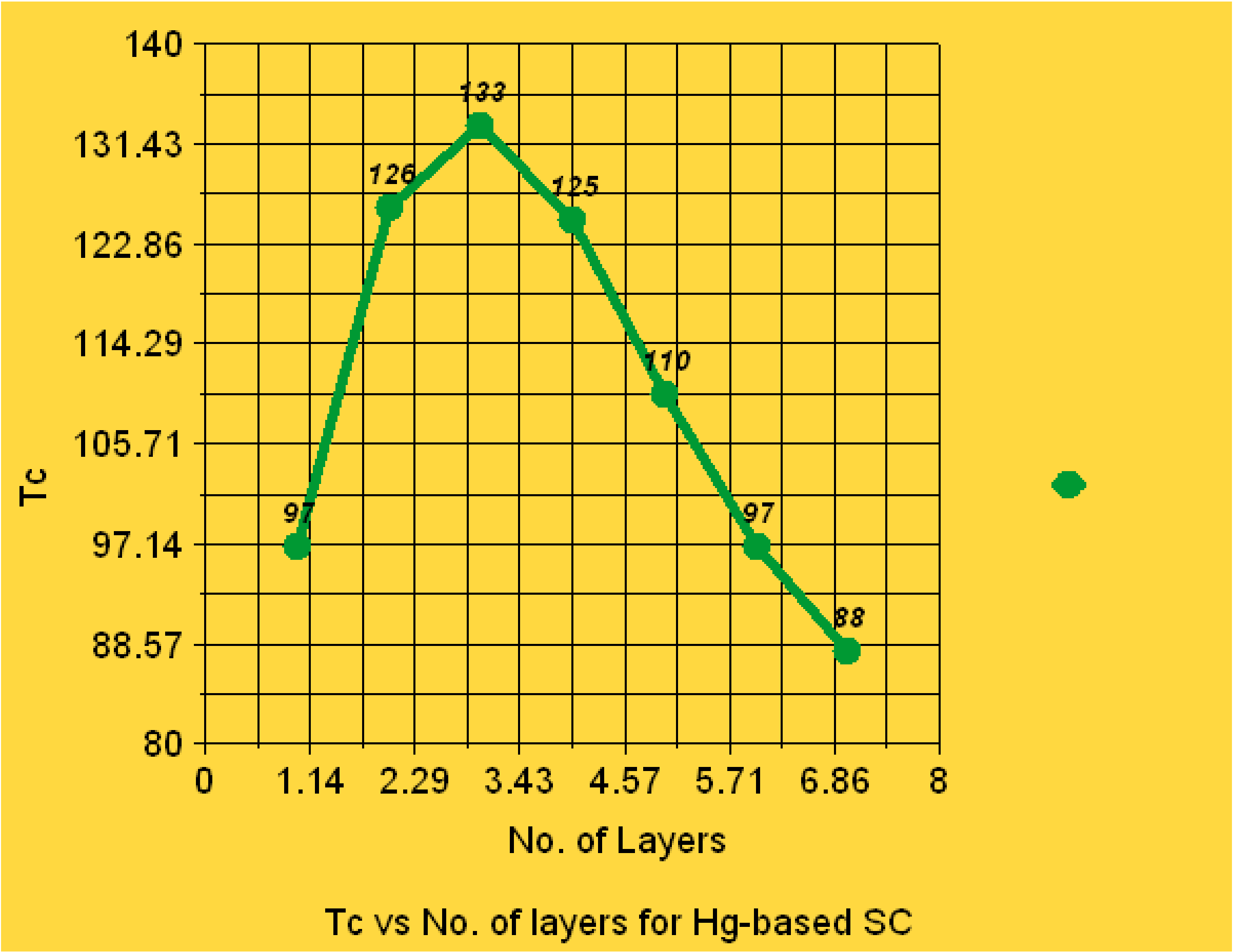

**Figure 6: Tc vs Number of layers for Mercury Cuprates. The graph shows that after attaining an optimum layer and peaking, the Tc then decreases as the layers increase This graph was first produced in references 32 and 33. Figure 5 bears a very close resemblance, suggesting a link between Fw/Z and layering in superconductivity.**